# ELEMENTARY ACT OF CONSCIOUSNESS OR "CYCLE OF MIND", INVOLVING DISTANT AND NONLOCAL INTERACTION

## Alex Kaivarainen
web.petrsu.ru/~alexk

Each macroscopic process can be subdivided on the quantum and classical stages.
*Particularly, the quantum stages of Cycle of Mind involves:*

1) the stimulation of dynamic correlation between water clusters in the same and remote microtubules (MTs) in state of mesoscopic Bose condensation (mBC) by phonons (acoustic waves) and by librational IR photons (electromagnetic-EM waves) distant exchange;

2) the transition from distant EM interaction between remote MTs to nonlocal quantum interaction, induced by IR photons exchange between clusters, the clusters Virtual Replicas multiplication (VRM) and virtual guides (VirG) formation between elementary particles of remote coherent water molecules. This process represents transition from mesoscopic Bose condensation to macroscopic nonuniform semivirtual Bose condensation (VirBC) (Kaivarainen, 2006, http://arxiv.org/abs/physics/0207027);

3) the collapsing of corresponding macroscopic wave function, as a result of the optical bistability of the entangled water clusters and their disassembly due to librational photons pumping, shifting clusters to less stable state;

4) turning the clusterphilic interaction between water clusters in the open state of cavities between alpha and beta tubulins to hydrophobic one and the in-phase shift of these cavities to the closed state due to clusters disassembly (Kaivarainen, http://arxiv.org/abs/physics/0102086).

*The classical stages of our Hierarchic model of elementary act of "Cycle of Mind" are following:*

a) the nerve cells membrane depolarization;

b) the gel → sol transition induced by disconnection of microtubules (MTs) with membranes and disassembly of the actin filaments;

c) the shape/volume pulsation of dendrites of big number of coherently interacting nerve cells, accompanied by jump-way reorganization of synaptic contacts on the dendrites surface;

d) the back sol → gel transition in corresponding cells ensembles, stabilizing (memorizing) of new state by formation of new system of microtubules and MTs associated proteins in the dendrites (MAPs).

The 'Period of Cycle of Mind' is determined by the sum of the life-time of quantum phase - semivirtual macroscopic Bose condensation, providing coherence and macroscopic entanglement and the life-time of collapsed mesoscopic Bose condensation, induced by decoherence factors. The life-time of entangled coherent phase can be many times shorter than classical phase. Consequently, the $[coherence \rightleftharpoons decoherence]$ dynamic equilibrium in macroscopic system of neurons is strongly shifted to the right. However, even extremely short time of macroscopic entanglement is enough for nonlocal remote interaction. This is true not only for systems under consideration, but for any kind of oscillating macroscopic entanglement.

Our approach to elementary act of consciousness has some common features with well-known Penrose - Hameroff model, interrelated act of consciousness with the wave function of microtubules collapsing. So we start from description of this Orchestrated objective reduction (Orch OR) model.

# 1  The basis of Orchestrated objective reduction (Orch OR) model of Penrose and Hameroff and comparison with Cycle of Mind model

Non-computable self-collapse of a quantum coherent wave function within the brain may fulfill the role of non-deterministic free will after Penrose and Hameroff (1995).

For biological qubits Penrose and Hameroff chose the open and closed clefts between the pair of tubulin subunits in microtubules. Tubulin qubits would interact and compute by entanglement with other tubulin qubits in microtubules in the same and different neurons.

It was known that the pair of alpha and beta tubulin subunits flexes 30 degrees, giving two different conformational shapes as superpositions.

The authors considered three possible types of tubulin superpositions: separation at the level of the entire protein, separation at the level of the atomic nuclei of the individual atoms within the proteins, and separation at the level of the protons and neutrons (nucleons) within the protein.

The calculated gravitational energy (E) at the level of atomic nuclei of tubulins had the highest energy, and would be the dominant factor in the wave function collapsing.

The best electrophysiological correlate of consciousness is gamma EEG, synchronized oscillations in the range of 30 to 90 Hz (also known as "coherent 40 Hz") mediated by dendritic membrane depolarizations (not axonal action potentials). This means that roughly 40 times per second (every 25 milliseconds) neuronal dendrites depolarize synchronously throughout wide regions of brain.

Using the indeterminacy principle $E = \hbar/t$ for OR, the authors take $t = 25$ ms, and calculated (E) in terms of number of tubulins (since $E$ was already evaluated for one tubulin).

The number of tubulins to be required in isolated superposition to reach OR threshold in $t = 25$ ms turned out to be $2 \times 10^{11}$ tubulins.

Each brain neuron is estimated to contain about $10^7$ tubulins (Yu and Bass, 1994). If 10% of each neuron's tubulins became coherent, then Orch OR of tubulins within roughly 20,000 (gap-junction connected) neurons would be required for a 25 ms conscious event, 5,000 neurons for a 100 ms event, or 1,000 neurons for a 500 ms event, etc.

These estimates (20,000 to 200,000 neurons) fit very well with others from more conventional approaches suggesting tens to hundreds of thousands of neurons are involved in consciousness at any one time.

The environmental decoherence can be avoided in quasi-solid (gelatinous: "gel") phases due to polymerization states of the actin. In the actin-polymerized gel phase, cell water and ions are ordered on actin surfaces, so microtubules are embedded in a highly structured (i.e. non-random) medium. Tubulins are also known to have C termini "tails", negatively charged peptide sequences extending string-like from the tubulin body into the cytoplasm, attracting positive ions and forming a plasma-like Debye layer which can also shield microtubule quantum states. Finally, tubulins in microtubules were suggested to be coherently pumped laser-like into quantum states by biochemical energy (as proposed by H. Fröhlich).

Actin state dependent gel⇌sol cycling occur with frequency 40 Hz. Thus during classical, liquid (sol) phases of actin depolymerization, inputs from membrane/synaptic inputs could "orchestrate" microtubule states. When actin gelation occurs, quantum isolation and computation ensues until OR threshold is reached, and actin depolymerizes.

The result of each OR event (in terms of patterns of tubulin states) would proceed to organize neuronal activities including axonal firing and synaptic modulation/learning. Each OR event (e.g. 40 per second) is proposed to be a conscious event.

One implication of the Orch OR model is that consciousness is a sequence of discrete events, related to collapsing of general for these states wave function.

However, the following problems are not clear in the Hamroff-Penrose approach:
a) what is the mechanism responsible for coherence of big number of remote tubulins and their entanglement;
b) how the microtubules can be unified by single wave function, like in the case of macroscopic Bose condensation, e.g. superfluidity or superconductivity?
*Our model of elementary act of consciousness or cycle of mind has the answers to this crucial questions.*

In our approach we explain the selection of certain configurational space of huge number of 'tuned' neurons, not by structural changes of tubulins like in Hameroff-Penrose model, but by increasing of mass of water in state of macroscopic BC in brain in the process of condensation of spatially separated mesoscopic BC - mBC (coherent water clusters in MTs), stimulated by IR photons exchange.

However, the macroscopic BC resulting from unification of mesoscopic BC is initiated by correlated shift of dynamic equilibrium of nonpolar clefts formed by tubulins, between the open (b) and closed (a) states to the open one, increasing the fracture of water clusters and their resulting mass. The corresponding structural rearrangements of tubulins pairs in the process of shift of [open $\rightleftharpoons$ closed] clefts to the right or left itself, do not change their mass. So, they can not be a source of wave function collapsing "under its own weight" in contrast to increasing of mass of water in state of entangled nonuniform semivirtual BC, proposed in our approach.

The dynamics of increasing $\rightleftharpoons$ decreasing of the entangled water mass in state of macroscopic BC is a result of correlated shift of dynamic equilibrium between primary librational (lb) effectons (coherent water clusters - mesoscopic BC), stabilized by open inter-tubulins cavities and primary translational (tr) effectons (independent water molecules), corresponding to closed cavities.

The correlated conversions between librational (lb) and translational (tr) effectons of water in remote MTs, representing the [*association* $\rightleftharpoons$ *dissociation*] of the entangled water clusters in state of mBC reflect, in fact, the reversible cycles of [coherence $\rightleftharpoons$ decoherence] corresponding to cycles of mesoscopic wave function of these clusters collapsing.

The relatively slow oscillations of dynamic equilibrium of $lb \rightleftharpoons tr$ conversions with period about $1/40 = 25$ ms are responsible for alternating contribution of macroscopic quantum entanglement and macroscopic wave function collapsing in human consciousness. *Each such reversible dynamic process represents the Elementary Act of Consciousness or "Cycle of Mind"*.

## 2. The basis of Elementary Act of Consciousness, including the distant and nonlocal interactions

In accordance to our model of elementary act of consciousness (EAC) or Cycle of Mind, each specific kind of neuronal ensembles excitation, accompanied by jump-way reorganization of big number of *dendrites and synaptic contacts* - corresponds to certain change of hierarchical system of three - dimensional (3D) standing waves of following kinds:
- thermal de Broglie waves (waves B), produced by anharmonic translations and librations of molecules;
- electromagnetic (IR) waves;
- acoustic waves;
- virtual pressure waves ($VPW^\pm$);
- inter-space waves (ISW), in-phase with $VPW^\pm$;

- virtual spin waves and (VirSW).

Scattering of all-penetrating virtual waves of Bivacuum, representing the *reference waves,* on the de Broglie waves of atoms and molecules of any mesoscopic or macroscopic object - generates the *object waves* by analogy with optical holography.

The interference of Bivacuum *reference waves* with the *object waves* forms the primary Virtual Replica (VR) of any systems, including: water clusters, microtubules (MTs) and whole neurons. The notion of primary VR of any material object was introduced earlier in theory of Bivacuum (Kaivarainen, 2006; 2007).

The *primary* **VR** of any macroscopic object represents 3D interference pattern of Bivacuum *reference waves*:
- virtual pressure waves (**VPW**$^{\pm}$),
- virtual spin waves (**VirSW**$^{\pm 1/2}$) and
- inter-space waves (**ISW**)

with the *object waves* of similar nature.

The resulting *primary* **VR** can be subdivided on the *surface* and *volume* virtual replicas: **VR**$_S$ and **VR**$_V$, correspondingly. The **VR**$_S$ contains the information about shape of the object, like regular optical hologram. The **VR**$_V$ contains the information about internal spatial and dynamic properties of the object.

The multiplication/iteration of the *primary* **VR** with properties of 3D standing waves in space and time was named Virtual Replica Multiplication: **VRM(r,t)**.

The **VRM(r,t)** can be named *Holoiteration* by analogy with **hologram**. In Greece *'holo'* means the 'whole' or 'total'.

The factors, responsible for conversion of mesoscopic Bose condensation (mBC) to macroscopic BC are following:

a) the IR photons exchange interaction between coherent water clusters in the same and remote microtubules (MTs),

b) the spatial Virtual Replica multiplication: **VRM(r)** and

c) formation of Virtual Guides of spin, momentum and energy (**VirG**$_{SME}$), connecting the entangled protons and neutrons of similar (like H$_2$O) remote coherent molecules and atoms (see Kaivarainen 2006; 2007).

*The nonuniform macroscopic BC can be considered as a unified sub-systems of connected by bundles of* **VirG**$_{SME}$ *nucleons of water clusters (actual mesoscopic BC), abridged by* **VRM(r,t)** *and* **VirG**$_{SME}$ *(virtual BC).*

The dynamics of [increasing ⇌ decreasing] of the entangled water mass in state of macroscopic BC in the process of elementary act of consciousness is a result of correlated shift of dynamic equilibrium between open and closed cavities formed by alpha and beta tubulins. The closed or open configuration of cavities, the relative orientation of microtubules and their stability as respect to disassembly can be regulated by variable system of microtubule - associated - proteins (MAPs).

The **Reason and Mechanism** of macroscopic Wave Function collapsing, induced by decoherence of VirBC can be following:

*The Reason* of collapsing is destabilization of macroscopic Bose condensation of coherent water clusters in state of mesoscopic BC in MTs.

*The Mechanism* is a consequence of such known quantum optic phenomena, as *bistability*. The bistability represents the $a \rightleftharpoons b$ equilibrium shift of librational primary effectons between the acoustic (a) and optic (b) state to the less stable (b) state, as a consequence of librational IR photons pumping and absorption by water clusters, saturating b-state. This saturation can be considered as the decoherence factor, triggering the macroscopic Wave Function collapsing.

The result of *bistability* is the collective dissociation of water clusters in MTs between tubulins and shift the nonpolar cavities states equilibrium to the closed state. This process is accompanied by shrinkage of MTs. *It is a transition from quantum to classical stages of "elementary act of consciousness".*

The shrinkage of MTs induces the disjoining of the MTs ends from the internal surfaces of membranes of nerve cell bodies. The consequence of disjoining is the [gel ⇌ sol] transition in cytoplasm, accompanied by disassembly of actin filaments.

Strong increasing of the actin monomers free surface and the fraction of water, involved in hydration shells of these proteins, decreases the internal water activity and initiate the passive osmosis of water into the nerve cell from the external space. The cell swallows and its volume increases. Corresponding change of cell's body volume and shape of dendrites is followed by synaptic contacts reorganization. *This is a final classical stage of act of consciousness or "Cycle of Mind".*

### 3. Mesoscopic Bose condensation (mBC) at physiological temperature. Why it is possible ?

The possibility of mesoscopic (intermediate between microscopic and macroscopic) Bose condensation in form of coherent molecular and atomic clusters in condensed matter (liquid and solid) at the ambient temperature was rejected for a long time. The reason of such shortcoming was a **wrong starting assumption**, that the thermal oscillations of atoms and molecules in condensed matter are harmonic ones (see for example: Beck and Eccles, 1992). The condition of harmonic oscillations means that the averaged kinetic ($T_k$) and potential ($V$) energy of molecules are equal to each other and linearly dependent on temperature ($T$):

$$T_k = V = \frac{1}{2}kT \qquad 1$$

where : $k$ is a Boltzmann constant

The averaged kinetic energy of the oscillating particle may be expressed via its averaged momentum ($p$) and mass ($m$):

$$T_k = p^2/2m \qquad 2$$

The most probable wave B length ($\lambda_B$) of such particle, based on wrong assumption (1), is:

$$\lambda_B = h/p = \frac{h}{(mkT)^{1/2}} \qquad 3$$

It is easy to calculate from this formula, that around the melting point of water $T = 273\,K$ the most probable wave B length of water molecule is about 1Å, i.e. much less than the distance between centers of $H_2O$ ($l \sim 3$ Å). This result leads to shortcoming that no mesoscopic Bose condensation (BC) is possible at this and higher temperature and water and ice are classical systems.

It is known from theory of Bose condensation that mesoscopic BC is possible only at conditions, when the length of waves B of particles exceeds the average distance between their centers ($l$):

$$l = (V_0/N_0)^{1/3} \qquad 4$$

$$L > \lambda_B > l \qquad 5$$

where: $L$ is a macroscopic parameter, determined by dimensions of the whole sample.

Condition (5) is a condition of partial or mesoscopic Bose condensation in form of coherent molecular clusters - named primary effectons, which confirmed to be correct in our Hierarchic theory of condensed matter and related computer calculations (see new book online: http://arxiv.org/abs/physics/0102086).

Consequently, the incorrect assumption (1) leads to formula (3) and incorrect result:

$$\lambda_B < l = (V_0/N_0)^{1/3} \qquad 6$$

meaning the absence of mesoscopic Bose condensation (mBC) in condensed matter at room and higher temperature.

The right way to proceed is to evaluate correctly the ratio between internal kinetic and potential energy of condensed matter and after this apply to Virial theorem (Clausius,1870).

It is shown (Kaivarainen, 1995; 2007), that the structural factor (S) of collective excitation, which can be calculated using our pCAMP computer program reflects the ratio of kinetic energy ($T_k$) to the total energy ($E = V + T_k$) of quasiparticle.

If $S = T_k/(V + T_k) < 1/2$ ($V > T_k$), this points to anharmonic oscillations of particles and nonclassical properties of corresponding matter in accordance to Virial theorem.

The Virial theorem in general form is correct not only for classical, but as well for quantum systems. It relates the averaged kinetic $\bar{T}_k(\vec{v}) = \sum_i \overline{m_i \mathbf{v}_i^2/2}$ and potential $\bar{V}(r)$ energies of particles, composing these systems:

$$2\bar{T}_k(\vec{v}) = \sum_i \overline{m_i \mathbf{v}_i^2} = \sum_i \overline{\vec{r}_i \partial V/\partial \vec{r}_i} \qquad 7$$

It follows from Virial theorem, that if the potential energy $V(r)$ is a homogeneous $n - order$ function:

$$V(r) \sim r^n \qquad 8$$

then the average kinetic and the average potential energies are related as:

$$n = \frac{2\overline{T_k}}{\overline{V(r)}} \qquad 9$$

For example, for a harmonic oscillator, when $\bar{T}_k = \bar{V}$, we have $n = 2$ and condition (1).
For Coulomb interaction: $n = -1$ and $\bar{T} = -\bar{V}/2$.
For water our hierarchic theory based computer calculations of $\bar{T}_k$ and $\bar{V}$ gives:
$n_w \sim 1/15$ $(\bar{V}/\bar{T}_k \sim 30)_w$ and for ice: $n_{\text{ice}} \sim 1/50$ $(\bar{V}/\bar{T}_k \sim 100)_{ice}$ (see http://arxiv.org/abs/physics/0102086).

It follows from (8) and our results, that in water and ice the dependence of potential energy on distance (r) is very weak:

$$V_w(r) \sim r^{(1/15)}; \qquad V_{\text{ice}} \sim r^{(1/50)} \qquad 10$$

Such weak dependence of potential energy on the distance can be considered as indication of long-range interaction due to the expressed cooperative properties of water as associative liquid and the ability of its molecules for mesoscopic Bose condensation

(mBC).

The difference between water and ice (10) and our computer simulations prove, that the role of distant Van der Waals interactions, stabilizing primary effectons (mesoscopic molecular Bose condensate), is increasing with temperature decreasing and [liquid→solid] phase transition. This correlates with strong jump of dimensions of **H$_2$O** clusters in state of mBC just after freezing, evaluated in our work.

The conditions (10) are good evidence that oscillations of molecules in water and ice are strongly anharmonic and the condensed matter in both phase can not be considered as a classical system. For real condensed matter we have:

$$\overline{T_k} \ll \overline{V} \quad \text{and} \quad \lambda_B > l = (V_0/N_0)^{1/3} \qquad 11$$

It is important to note, that when the average momentum ($\overline{p}$) is tending to zero for example with temperature decreasing or pressure increasing, the kinetic energy is also tending to zero:

$$\overline{T_k} = \overline{p}^2/2m \; \to \; 0 \qquad 12$$

then the ratio: $n = \frac{2\overline{T_k}}{\overline{V(r)}} \to 0$ and the interaction between particles becomes independent on the distance between them:

$$V_w(r) \sim r^{(n \to 0)} = 1 = const \qquad 13$$

*Consequently the interaction turns from the regular distant interaction to the nonlocal one.*

This transition is in-phase with turning of the particles system from state of mesoscopic Bose condensation to macroscopic one. We came to important conclusion, that the conditions of nonlocality (12 and 13) become valid at conditions of macroscopic BC at $\overline{p} \to 0$:

$$\overline{\lambda}_B = (h/\overline{p}) \; \to \; \infty \qquad 14$$

This macroscopic BC can be *actual*, like in the case superconductivity and superfluidity, it can be virtual BC, composed from Bivacuum dipoles (see http://arxiv.org/abs/physics/0207027) and it can be *semiactual or semivirtual BC*. The latter will be discussed in the next section.

### 4. The transition from distant electromagnetic interaction between remote water clusters (mBC) to macroscopic Bose condensation and nonlocal interaction

The role of Virtual Replica of clusters in state of mBC spatial multiplication VRM(r) (see http://arxiv.org/abs/physics/0207027) is to create the virtual connections between remote actual clusters. The subsequent formation of Virtual Channels between pairs of coherent elementary particles (electrons, protons and neutrons) of opposite spins, - turns the mesoscopic BC to macroscopic one. This is a result of unification of remote clusters wave function to integer linear superposition of its eigenvalues, resulting in macroscopic wave function. Part of these eigenvalues of the integer wave function corresponds to virtual replica (3D virtual standing waves) of the cluster and other to clusters themselves. The corresponding nonuniform macroscopic Bose condensate, consequently, is partly virtual and partly actual.

The transition from distant EM interaction between remote MTs to nonlocal quantum interaction is a result of entanglement between clusters - mBC, stimulated by IR photons exchange and Virtual Replicas multiplication (VRM) of the clusters. This transition is accompanied by nonuniform macroscopic virtual Bose condensation (VirBC). *The nonuniform VirBC become possible only at certain spatial separation and orientation of coherent water clusters as 3D standing de Broglie waves of water molecules in the entangled microtubules.*

The mechanism of this transition is based on our theories of Virtual Replica multiplication in space - VRM(r), virtual Bose condensation (VirBC) and nonlocal virtual guides (VirG$_{S,M,E}$) of spin, momentum and energy (Kaivarainen, 2006; 2007; http://arxiv.org/abs/physics/0207027).

The Virtual Guides have a shape of virtual microtubules with properties of quasi-1-dimensional virtual Bose condensate (Fig.1). The VirG are constructed from 'head-to-tail' polymerized Bivacuum bosons $BVB^{\pm}$ or Cooper pairs of Bivacuum fermions $[BVF^{\uparrow} \bowtie BVF^{\downarrow}]$. The bundles of VirG$_{S,M,E}$, connecting coherent atoms of Sender (S) and Receiver (S), named the *Entanglement Channels*, are responsible for macroscopic entanglement, providing nonlocal interaction, telepathy, remote healing and telekinesis. The poltergeist can be considered as a private case of telekinesis, realized via Entanglement Channels, connecting coherent elementary particles of psychic and the object.

The changes of de Broglie waves of atoms and molecules, participating in elementary act of consciousness, modulate virtual pressure waves of Bivacuum (**VPW**$^{\pm}$). These modulated standing virtual waves form the *Virtual Replica (VR)* of 'tuned' neuronal ensembles and of microtubules (MTs) systems.

*The notion of Virtual Replica of any material object* was introduced by this author in Unified theory of Bivacuum, duality of particles, time and fields (Kaivarainen, 2006; 2007). It is subdivided on the *surface and the volume Virtual Replicas.*

*The surface VR* have the resemblance with regular optical hologram and contains information only about 3D shape of the object. *The volume VR* is a consequence of penetration of Virtual Pressure Waves throw the object and scattering on its internal de Broglie waves of particles. So the volume VR contains info about the dynamic interior of the object. Consequently, the resulting VR of the object is much more informative than the regular optical hologram.

The classical stages Cycle of Mind represents the gel-sol transition in cytoplasm of the neuron bodies and the tuned neurons ensembles in-phase pulsation and axonal firing, accompanied by redistribution of synaptic contacts between the starting and final states of dendrites.

The quantum stage of Cycle Mind involves the collapsing of unified wave function of big number of the entangled coherent water clusters (mBC) in remote microtubules. This means a transition from the state of non-uniform semivirtual macroscopic Bose condensation to mesoscopic BC. It represents the reversible dissipative process.

The Hameroff - Penrose model considers only the coherent conformational transition of cavities between big number of pairs of tubulins between open to closed states in remote entangled MTs as the act of wave function collapsing.

In our model this transition is only the triggering act, stimulating quantum transition of the big number of entangled water clusters in state of coherent macroscopic Bose condensation to state of mesoscopic Bose condensation (mBC).

*Consequently, the Cycles of Mind can be considered as a reversible transitions of certain part of brain between coherence and decoherence, involving quantum and classical stages.* The non-uniform coherent state of semivirtual macroscopic Bose Condensation of

water in microtubules system in time and space, e.g. periodically entangled, but spatially separated flickering clusters of mBC is different of continuous macroscopic Bose Condensation, pertinent for superfluidity and superconductivity.

The Penrose-Hameroff model considers only the correlated conformational transition of cavities between pairs of tubulins between open to closed states of big number of MTs as the act of wave function collapsing. In our model this transition is only the triggering act, stimulating quantum transition of the big number of entangled water clusters in state of coherent macroscopic Bose condensation to state of nonentangled mesoscopic Bose condensation (mBC). Consequently, the cycles of consciousness can be considered as a reversible transitions of certain part of brain between *coherence* and *decoherence*.

The most important collective excitations providing the entanglement and quantum background of consciousness and can be the quantum integrity of the whole organism are coherent water clusters (primary librational effectons), representing mBC of water in microtubules.

Due to rigid core of MTs and stabilization of water librations (decreasing of most probable librational velocity $\mathbf{v}_{lb}$) the dimensions of mBC inside the MTs ($\lambda_B^{lb} = h/m\mathbf{v}_{lb}$) are bigger, than in bulk water and cytoplasm. The dimensions and stability of mBC is dependent on relative position of nonpolar cavities between $\alpha$ and $\beta$ tubulins, forming MTs and cavities dynamics.

In the open state of cavities the water clusters (mBC) are assembled and stable, making possible the macroscopic BC via entanglement in a big number of neurons MTs and in closed state of protein cavities the clusters are disassembled and macroscopic entanglement is destroyed.

The quantum beats between the ground - acoustic (a) and excited - optic (b) states of primary librational effectons (mBC) of water are accompanied by super-radiation of coherent librational IR photons and their absorption (see Introduction and Kaivarainen, 1992). The similar idea for water in microtubules was proposed later by Jibu at al. (1994, p.199).

The process of coherent IR photons radiation ⇌ absorption is interrelated with dynamic equilibrium between open (B) and closed (A) states of nonpolar clefts between $\alpha$ and $\beta$ tubulins. These IR photons exchange interaction between 'tuned' systems of MTs stands for *distant* interaction between neurons in contrast to nonlocal interaction provided by conversion of mesoscopic BC to macroscopic BC.

The collective shift in geometry of nonpolar clefts/pockets equilibrium from the open to closed state is accompanied by the shrinkage of MTs is a result of turning of clusterphilic interaction to hydrophobic ones and *dissociation of water clusters*. This process induce the disjoining of the MTs ends from the membranes of nerve cell bodies and gel → sol transition in cytoplasm, accompanied by disassembly of actin filaments.

Strong abrupt increasing of the actin monomers free surface and the fraction of water, involved in hydration shells of these proteins, decreases the internal water activity and initiate the water passive osmosis into the nerve cell from the external space. The cell swallows and its volume increases. Corresponding change of cell's body volume and shape of dendrites is followed by synaptic contacts reorganization. This is a final stage of multistage act of consciousness.

The *bistability* represents the water clusters polarization change as a result of $a \rightleftharpoons b$ equilibrium shift in librational primary effectons to the right. In turn, this shift is a consequence of librational IR photons pumping and the excited b-state of librational effectons saturation.

The related to above phenomena: the *self-induced transparency* is due to light absorption saturation by primary librational effectons (Andreev, et al., 1988). This saturation can be followed by the pike regime (light emission pulsation, after $b$−state saturation of librational effectons and subsequent super-radiation of big number of entangled water clusters in state of mBC in the process of their correlated $\sum(\mathbf{b} \to \mathbf{a})$ transitions.

The entanglement between coherent nucleons of opposite spins of H and O of remote water clusters in a big number of MTs, in accordance to our theory of nonlocality, can be realized via bundles of Bivacuum virtual guides **VirG**$_{SME}$ of spin, momentum and energy (Kaivarainen, 2006, 2007).

### 4.1 The mechanism of the Entanglement channels formation between remote coherent de Broglie waves of the nucleons

The bundles of **VirG**$_{SME}$, connecting pairs of protons and neutrons of opposite spins of remote coherent molecules in state of mesoscopic Bose condensation (mBC) were named the *Entanglement channels* (Kaivarainen, 2006, 2007):

$$\textbf{Entanglement channel} = \left[ \mathbf{N(t,r)} \times \sum_{}^{\mathbf{n}} \mathbf{VirG}_{SME}(\mathbf{S} <=> \mathbf{R}) \right]_{x,y,z}^{i} \qquad 15$$

where: ($\mathbf{n}$) is a number of pairs of similar tuned elementary particles (protons, neutrons and electrons) of opposite spins of the remote entangled atoms and molecules; $N(t,r)$ is a number of coherent atoms/molecules in the entangled molecular (e.g. water) clusters in state of mBC.

The Virtual Guides (microtubules), connecting the remote elementary particles have a properties of quasi- one- dimensional virtual Bose condensate.

A *single* Virtual Guides of spin, momentum and energy (see Fig. 1) are assembled from 'head-to-tail' polymerized Bivacuum bosons of opposite polarization: $\mathbf{BVB}^+ = [\mathbf{V}^+ \uparrow\downarrow \mathbf{V}^-]$ and $\mathbf{BVB}^- = [\mathbf{V}^+ \downarrow\uparrow \mathbf{V}^-]$ :

$$\textit{Single } \mathbf{VirG}_{SME}^{\mathbf{BVB}^+} = \mathbf{D(r,t)} \times \mathbf{BVB}^+; \qquad \mathbf{VirG}_{SME}^{\mathbf{BVB}^-} = \mathbf{D(r,t)} \times \mathbf{BVB}^- \qquad 16$$

A *double* Virtual Guides are composed from Cooper pairs of Bivacuum fermions of opposite spins ($\mathbf{BVF}^\uparrow \bowtie \mathbf{BVF}^\downarrow$):

$$\textit{Double } \mathbf{VirG}_{SME}^{\mathbf{BVF}^\uparrow \bowtie \mathbf{BVF}^\downarrow} = \mathbf{D(r,t)} \times [\mathbf{BVF}_+^\uparrow \bowtie \mathbf{BVF}_-^\downarrow]_{S=0}^{s} \qquad 17$$

where: $\mathbf{D(r,t)}$ is a number of Bivacuum dipoles in Virtual guides, dependent on the distance ($\mathbf{r}$) between remote but tuned de Broglie waves of elementary particles of opposite spins. The diameter of these dipoles and spatial gap between their torus and antitorus are pulsing in-phase.

Just the *Entanglement channels* are responsible for nonlocal Bivacuum mediated interaction between the mesoscopic BC, turning them to macroscopic BC. For the entanglement channels activation the interacting mBC systems should be in non-equilibrium state.

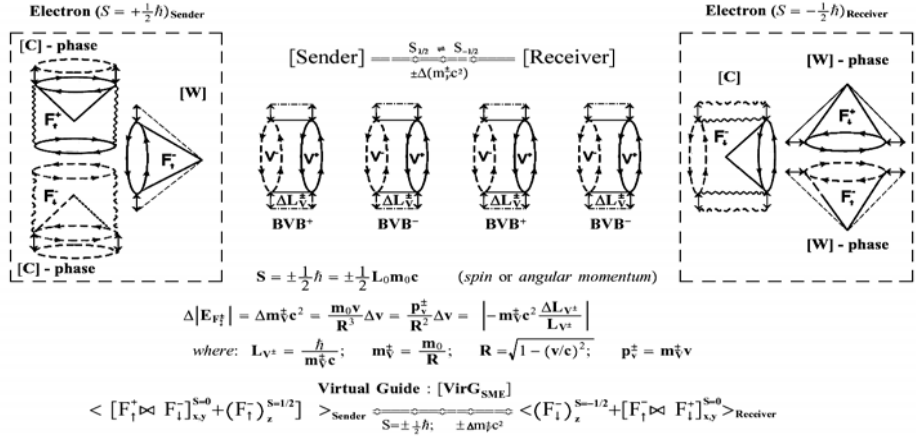

**Figure 1**. The mechanism of nonlocal Bivacuum mediated interaction (entanglement) between two distant unpaired sub-elementary fermions of 'tuned' elementary triplets (particles) of the opposite spins $< [\mathbf{F}_\uparrow^+ \bowtie \mathbf{F}_\downarrow^-] + \mathbf{F}_\uparrow^- >_{Sender}^i$ and $< [\mathbf{F}_\downarrow^+ \bowtie \mathbf{F}_\uparrow^-] + \mathbf{F}_\downarrow^- >_{Receiver}^i$, with close frequency of $[\mathbf{C} \rightleftharpoons \mathbf{W}]$ pulsation and close de Broglie wave length ($\lambda_B = \mathbf{h}/\mathbf{m}_V^+ \mathbf{v}$) of particles. The tunnelling of momentum and energy increments: $\Delta|\mathbf{m}_V^\pm \mathbf{c}^2| \sim \Delta|\mathbf{VirP}^+| \pm \Delta|\mathbf{VirP}^-|$ from Sender to Receiver and vice-verse via Virtual spin-momentum-energy Guide [$\mathbf{VirG}_{SME}^i$] is accompanied by instantaneous pulsation of diameter ($2\Delta \mathbf{L}_V^\pm$) of this virtual guide, formed by Bivacuum bosons $\mathbf{BVB}^\pm$ or double microtubule, formed by Cooper pairs of Bivacuum fermions: [$\mathbf{BVF}^\uparrow \bowtie \mathbf{BVF}^\downarrow$]. The nonlocal spin state exchange between [S] and [R] can be induced by the change of polarization of Cooper pairs: [$\mathbf{BVF}^\uparrow \bowtie \mathbf{BVF}^\downarrow$] $\rightleftharpoons$ [$\mathbf{BVF}^\downarrow \bowtie \mathbf{BVF}^\uparrow$] and Bivacuum bosons: $\mathbf{BVB}^+ \rightleftharpoons \mathbf{BVB}^-$, composing the double or single $\mathbf{VirG}_{SME}(\mathbf{S} <=> \mathbf{R})^i$, correspondingly (see Kaivarainen, 2006; 2007).

The assembly of huge number of bundles of virtual microtubules of Bivacuum, like Virtual Channels side-by-side can compose virtual multilayer membranes. Each of this layer, pulsing in counterphase with the next one between the excited and ground states are interacting with each other via dynamic exchange by pairs of virtual pressure waves [$\mathbf{VPW}^+ \bowtie \mathbf{VPW}^-$]. This process occur without violation of the energy conservation law and is accompanied by nonlocal Bivacuum gap oscillation over the space of virtual BC of Bivacuum dipoles. The value of spatial gap between the actual and complementary torus and antitorus of Bivacuum fermions is dependent on their excitation state quantum number ($\mathbf{n} = \mathbf{0}, \mathbf{1}, \mathbf{2}, \mathbf{3}\ldots$):

$$[\mathbf{d}_{V^+ \Updownarrow V^-}]_n = \frac{h}{\mathbf{m}_0 \mathbf{c}(\mathbf{1} + \mathbf{2n})} \qquad 18$$

The Bivacuum gap oscillations and corresponding inter-space waves (ISW), correlated with $\mathbf{VPW}^\pm$, can be responsible for the lateral or transversal nonlocality of Bivacuum in contrast to longitudinal one, connecting the nucleons with opposite spin, realized via $\mathbf{VirG}_{SME}$ (Kaivarainen, 2006a, b).

The gel-sol transition in the number of entangled neurons is accompanied by decreasing of viscosity of cytoplasm. The tuned - parallel orientation of MTs in tuned remote cells change as a result of Brownian motion, accompanied by decoherence and loosing the entanglement between water clusters in MTs.

This is followed by relaxation of the internal water + microtubulins to normal dynamics and grows of (+) ends of MTs up to new contacts formation with cells membranes, stabilizing cells dendrites new geometry and synaptic contacts distribution. *This new configuration and state of the nerve system and brain, represents the transition of the Virtual Replica of the brain from the former state to the new one.*

We assume in our model the existence of back reaction between the properties of Virtual Replica of the nerve system of living organisms, with individual properties, generated by systems:

$$[microtubules\ of\ neurons + DNA\ of\ chromosomes]$$

and the actual object - the organism itself. The corresponding subsystems can be entangled with each other by the described above Virtual Channels.

The interference of such individual (self) virtual replica VR{self} with virtual replicas of other organisms and inorganic macroscopic system may modulate the properties of VR{self}.

Because of back reaction of VR{self} on corresponding organism, the interaction of this organism with resulting/Global virtual replica of the external macroscopic world may be realized.

The twisting of centrioles in cells to parallel orientation, corresponding to maximum energy of the MTs interaction of remote neurons, is a first stage of the *next* elementary act of consciousness.

The superradiated photons from enlarged in MTs water clusters have a higher frequency than $(\mathbf{a} \rightleftharpoons \mathbf{b})_{lb}$ transitions of water primary librational effectons of cytoplasmic and inter-cell water. This feature provides the regular *transparency* of medium between 'tuned' microtubules of remote cells for librational photons.

The [gel→sol] transitions in cells is interrelated with tuned nerve cells (ensembles) coherent excitation, their membranes depolarization and the *axonal firing*.

### 5. Two triggering mechanisms of elementary act of consciousness

It is possible in some cases, that the excitation/depolarization of the nerve cells by the external factors (sound, vision, smell, tactical feeling) are the triggering - *primary* events and [gel→sol] transitions in nerve cells are the *secondary* events.

However, the opposite mechanism, when the tuning of remote cells and [gel → sol] transitions are the *primary* events, for example, as a result of thinking/meditation and the nerve cells depolarization of cells are *secondary* events, is possible also.

The 1st mechanism, describing the case, when depolarization of nerve membranes due to external factors is a *primary* event and *gel → sol* transition a *secondary* one, includes the following stages of elementary act of consciousness:

a) simultaneous depolarization of big enough number of neurons, forming ensemble, accompanied by opening the potential-dependent channels and increasing the concentration of $Ca^{2+}$ in cytoplasm of neurons body;

b) collective disassembly of actin filaments, accompanied by [gel → sol] transition of big group of depolarized neurons stimulated by $Ca^{2+}$ − activated proteins like gelsolin and villin. Before depolarization the concentration of $Ca^{2+}$ outside of cell is about $10^{-3}M$ and inside about $10^{-7}M$. Such strong gradient provide fast increasing of these ions concentration in cell till $10^{-5}M$ after depolarization.

c) strong decreasing of cytoplasm viscosity and disjoining of the (+) ends of MTs from membranes, making possible the spatial fluctuations of MTs orientations inducing decoherence and switching off the entanglement between mBC;

d) volume/shape pulsation of neuron's body and dendrites, inducing reorganization of ionic channels activity and synaptic contacts in the excited neuron ensembles. These volume/shape pulsations occur due to reversible decrease of the intra-cell water activity and corresponding swallow of cell as a result of increasing of passive osmotic diffusion of water from the external space into the cell.

In the opposite case, accompanied the process of braining, the depolarization of nerve membranes, the axonal firing is a *secondary* event and *gel* → *sol* transition a *primary* one, stimulated in turn by simultaneous dissociation of big number of water clusters to independent molecules. The latter process represents the conversion of primary librational effectons to translational ones, following from our theory (see 'convertons' in the Introduction of book: Kaivarainen, 2007, http://arxiv.org/abs/physics/0102086).

The frequency of electromagnetic field, related to change of ionic flux in excitable tissues usually does not exceed $10^3$ Hz (Kneppo and Titomir, 1989).

The electrical recording of human brain activity demonstrate a coherent (40 to 70 Hz) firing among widely distributed and distant brain neurons (Singer, 1993). Such synchronization in a big population of groups of cells points to possibility of not only the regular axon-mediated interaction, but also to fields-mediated interaction and quantum entanglement between remote neurons bodies.

The dynamic virtual replicas (VR) of all hierarchical sub-systems of brain and its space-time multiplication VRM(r,t) contain information about all kind of processes in condensed matter on the level of coherent elementary particles (Kaivarainen, 2006 a,b). Consequently, our model agrees in some points with ideas of Karl Pribram (1977), David Bohm and Basil Hiley (1993) of holographic mind, incorporated in the hologram of the Universe.

### 6. The comparison of Hierarchic model of consciousness and Quantum brain dynamics model

Our approach to Quantum Mind problem has some common features with model of Quantum Brain Dynamics (QBD), proposed by L.Riccardi and H.Umezawa in 1967 and developed by C.I.Stuart, Y.Takahashi, H.Umezava (1978, 1979), M.Jibu and K.Yasue (1992, 1995).

In addition to traditional electrical and chemical factors in the nerve tissue function, this group introduced two new types of *quantum* excitations (ingredients), responsible for the overall control of electrical and chemical signal transfer: *corticons and exchange bosons* (dipolar phonons).

The *corticons* has a definite spatial localization and can be described by Pauli spin matrices. The *exchange bosons*, like phonons are delocalized and follow Bose-Einstein statistics. "By absorbing and emitting bosons coherently, corticons manifest global collective dynamics…, providing systematized brain functioning" (Jibu and Yasue,1993). In other paper (1992) these authors gave more concrete definitions:

"*Corticons* are nothing but quanta of the molecular vibrational field of biomolecules (quanta of electric polarization, confined in protein filaments). *Exchange bosons* are nothing but quanta of the vibrational field of water molecules…".

It is easy to find analogy between spatially localized "corticons" and our primary effectons as well as between "exchange bosons" and our secondary (acoustic) deformons. It is evident also, that our Hierarchic theory is more developed as far as it is based on detailed description of all collective excitations in any condensed matter (including water and biosystems) and their quantitative analysis.

Jibu, Yasue, Hagan and others (1994) discussed a possible role of quantum optical coherence in microtubules for brain function. They considered MTs as a *wave guides* for coherent superradiation. They also supposed that coherent photons, penetrating in MTs, lead to "*self-induced transparency*". Both of these phenomena are well known in fiber and quantum optics. We also use these phenomena for explanation of transition from mesoscopic entanglement of water clusters in MTs to macroscopic one, as a result of IR photons exchange between coherent clusters (mBC). However, we have to note, that the transition of mBC to macroscopic BC in 'tuned' MTs is possible without self-induced transparency also.

It follows also from our approach that the mechanism of macroscopic BC of water clusters do not need the hypothesis of Frölich that the proteins (tubulins of MTs) can be coherently pumped into macroscopic quantum states by biochemical energy.

We also do not use the idea of Jibu et al. that the MTs works like the photons wave - guides without possibility of side radiation throw the walls of MTs. The latter in our approach increases the probability of macroscopic entanglement between remote MTs and cells of the organism.

### 7. The Properties of the Actin Filaments, Microtubules and Internal Water

There are six main kind of actin existing. Most general F-actin is a polymer, constructed from globular protein G-actin with molecular mass 41800. Each G-actin subunit is stabilized by one ion $Ca^{2+}$ and is in noncovalent complex with one ATP molecule. Polymerization of G-actin is accompanied by splitting of the last phosphate group. The velocity of F-actin polymerization is enhanced strongly by hydrolysis of ATP. However, polymerization itself do not needs energy. Simple increasing of salt concentration (decreasing of water activity), approximately till to physiological one - induce polymerization and strong increasing of viscosity.

The actin filaments are composed from two chains of G-actin with diameter of 40 Å and forming double helix. The actin filaments are the polar structure with different properties of two ends.

*Let us consider the properties of microtubules (MT) as one of the most important component of cytoskeleton, responsible for spatial organization and dynamic behavior of the cells.*

The microtubules (MTs) are the nanostructures of cells, interconnecting the quantum and classical stages of the Cycle of Mind.

The [assembly ⇔ disassembly] equilibrium of microtubules composed of *α* and *β* tubulins is strongly dependent on internal and external water activity (*a*), concentration of $Ca^{2+}$ and on the electric field gradient change due to MTs piezoelectric properties.

The *α* and *β* tubulins are globular proteins with equal molecular mass ($MM = 55.000$), usually forming *αβ* dimers with linear dimension 8 nm. Polymerization of microtubules can be stimulated by NaCl, $Mg^{2+}$ and GTP (1:1 tubulin monomer) (Alberts *et al.*, 1983). The presence of heavy water (deuterium oxide) also stimulates polymerization of MTs.

In contrast to that the presence of ions of $Ca^{2+}$ even in micromolar concentrations, action of colhicine and lowering the temperature till $4^0C$ induce disassembly of MT.

Due to multigenic composition, *α* and *β* tubulins have a number of isoforms. For example, two-dimensional gel electrophoresis revealed 17 varieties of *β* tubulin in mammalian brain (Lee *et al.*, 1986). Tubulin structure may also be altered by enzymatic modification: addition or removal of amino acids, glycosylation, etc.

*Microtubules* are hollow cylinders, filled with water. Their internal diameter about $d_{in} = 140$Å and external diameter $d_{ext} = 280$Å (Figure 2). These data, including the dimensions of *αβ* dimers were obtained from x-ray crystallography (Amos and Klug, 1974). However we must keep in mind that under the conditions of crystallization the multiglobular proteins and their assemblies tends to more compact structure than in solutions due to lower water activity.

This means that in natural conditions the above dimensions could be a bit bigger.

The length of microtubules (MT) can vary in the interval:

$$l_t = (1 - 20) \times 10^5 Å \qquad 19$$

The spacing between the tubulin monomers in MT is about 40 Å and that between $\alpha\beta$ dimers: 80 Å are the same in longitudinal and transversal directions of MT.

Microtubules sometimes can be as long as axons of nerve cells, *i.e.* tenth of centimeters long. Microtubules (MT) in axons are usually parallel and are arranged in bundles. Microtubules associated proteins (MAP) form a "bridges", linking MT and are responsible for their interaction and cooperative system formation. Brain contains a big amount of microtubules. *Their most probable length is about $10^5 Å$, i.e. close to librational photon wave length.*

The viscosity of ordered water in such narrow microtubules seems to be too high for transport of ions or metabolites at normal conditions.

All 24 types of quasi-particles, introduced in the Hierarchic Theory of matter (Table 1), also can be pertinent for ordered water in the microtubules (MT). However, the dynamic equilibrium between populations of different quasi-particles of water in MT must be shifted towards primary librational effectons, comparing to bulk water due to increased clusterphilic interactions (Kaivarainen, 1985, 2000, 2007). The dimensions of internal primary librational effectons have to be bigger than in bulk water as a consequence of stabilization of MT walls the mobility of water molecules, increasing their most probable de Broglie wave length.

The interrelation must exist between properties of internal water in MT and structure and dynamics of their walls, depending on $[\alpha - \beta]$ tubulins interaction. Especially important can be a quantum transitions like convertons $[tr \Leftrightarrow lb]$. The convertons in are accompanied by [dissociation/association] of primary librational effectons, i.e. flickering of coherent water clusters, followed by the change of angle between $\alpha$ and $\beta$ subunits in tubulin dimers.

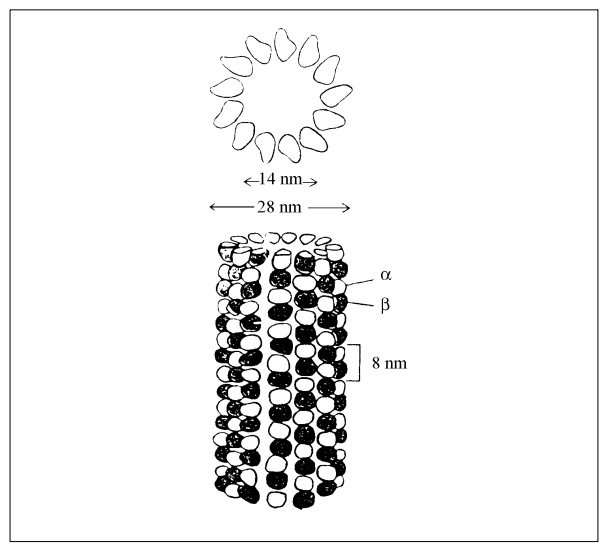

**Figure 2**. Construction of microtubule from $\alpha$ and $\beta$ tubulins, globular proteins with molecular mass 55 kD, existing in form of dimers ($\alpha\beta$). Each $\alpha\beta$ dimer is a dipole with negative charges, shifted towards $\alpha$ subunit (De Brabander, 1982). Consequently, microtubules, as an oriented elongated structure of dipoles system, have the piezoelectric

properties (Athestaedt, 1974; Mascarennas, 1974).

Intra-microtubular *clusterphilic interactions* stimulate the growth of tubules from $\alpha\beta$ tubulin dimers. The structural physical-chemical asymmetry of $\alpha\beta$ dimers composing microtubules determines their different rates of growth from the opposite ends ([+] and [-]).

The equilibrium of "closed" (A) and "open"(B) states of nonpolar cavities between $\alpha$ and $\beta$ tubulins in ($\alpha\beta$) dimers can be shifted to the (B) one under the change of external electric field in a course of membrane depolarization. It is a consequence of piezoelectric properties of MTs and stimulate the formation of coherent water clusters in the open cavities of ($\alpha\beta$) dimers. The open cavities serve as a centers of water cluster formation and molecular Bose condensation.

The parallel orientation of MT in different cells, optimal for maximum [MT-MT] resonance interaction could be achieved due to twisting of centrioles, changing spatial orientation of MT. However, it looks that the normal orientation of MT as respect to each other corresponds to the most stable condition, *i.e.* minimum of potential energy of interaction (see Albreht-Buehner, 1990).

It is important to stress here that the orientation of two centrioles as a source of MT bundles in each cell are always normal to each other.

The linear dimensions of the primary librational effectons edge ($l_{ef}^{lb}$) in pure water at physiological temperature ($36^0C$) is about 11 Å and in the ice at $0^0C$ it is equal to 45 Å.

We assume that in the rigid internal core of MT, the linear dimension (edge length) of librational effecton, approximated by cube is between 11Å and 45 Å *i.e.* about $l_{ef}^{lb} \sim 23$Å. It will be shown below, that this assumption fits the spatial and symmetry properties of MT very well.

The most probable group velocity of water molecules forming primary *lb* effectons is:

$$\mathbf{v}_{gr}^{lb} \sim h/(m_{H_2O} \times l_{ef}^{lb}) \qquad 20.$$

The librational mobility of internal water molecules in MT, which determines ($\mathbf{v}_{gr}^{lb}$) should be about 2 times less than in bulk water at $37^0C$, if we assume for water in microtubules: $l_{ef}^{lb} \sim 23$Å.

Results of our computer simulations for pure bulk water shows, that the distance between centers of primary [lb] effectons, approximated by cube exceed their linear dimension to about 3.5 times (Fig 3 b). For our case it means that the average distance between the effectons centers is about:

$$d = l_{ef}^{lb} \times 3.5 = 23 \times 3.5 \sim 80\text{Å} \qquad 21$$

This result of our theory points to the equidistant (80 Å) localization of the primary *lb* effectons in clefts between $\alpha$ and $\beta$ tubulins of each ($\alpha\beta$) dimer in the internal core of MTs.

In the case, if the dimensions of librational effectons in MTs are quite the same as in bulk water, i.e. 11 Å, the separation between them should be:
$d = l_{ef}^{lb} \times 3.5 = 11 \times 3.5 \sim 40$ Å.

This result points that the coherent water clusters can naturally exist not only between $\alpha$ and $\beta$ subunits of each pair, but also between pairs of ($\alpha\beta$) dimers.

In the both cases the spatial distribution symmetry of the internal flickering clusters in MT (Fig 2; 3) may serve as an important factor for realization of the signal propagation along the MT (conformational wave), accompanied by alternating process of closing and opening the clefts between neighboring $\alpha$ and $\beta$ tubulins pairs.

This large-scale protein dynamics is regulated by dissociation $\rightleftharpoons$ association of water clusters in the clefts between ($\alpha\beta$) dimers of MT (Fig.2) due to [*lb/tr*] convertons excitation and librational photons and phonons exchange between primary and secondary effectons, correspondingly.

The dynamic equilibrium between *tr* and *lb* types of the intra MT water effectons must to be very sensitive to $\alpha - \beta$ tubulins interactions, dependent on nerve cells excitation and their membranes polarization.

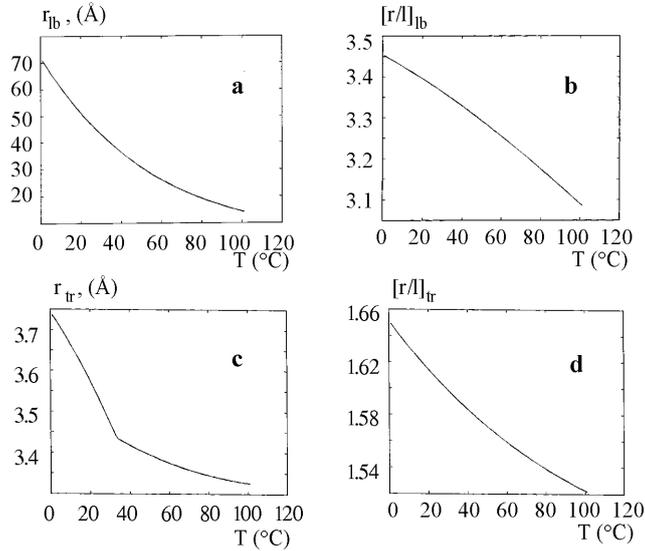

**Figure 3**. Theoretical temperature dependencies of:

(a) - the space between centers of primary [lb] effectons (calculated in accordance to eq.4.62 of http://arxiv.org/abs/physics/0102086);

(b) - the ratio of space between primary [lb] effectons to their length (calculated, using eq.4.63 of http://arxiv.org/abs/physics/0102086);

(c) - the space between centers of primary [tr] effectons (in accordance to eq.4.62);

(d) - the ratio of space between primary [tr] effectons to their length (eq.4.63).

**Two statements of our Hierarchic model of consciousness are important**:

1. The ability of intra-MT primary water effectons (tr and lb) for superradiation of six coherent IR photons from each of the effectons side, approximated by parallelepiped:

two identical - "longitudinal" IR photons, penetrating along the core of microtubule, forming the longitudinal standing waves inside it and two pairs of identical - "transverse" IR photons, also responsible for the distant, nonlocal interaction between microtubules. *In accordance to superradiation mechanism the intensity of longitudinal radiation of MTs is much bigger than that of transverse one;*

2. The parameters of the water clusters radiation (frequency of librational photons, coherency, intensity) are regulated by the interaction of the internal water with MT walls, dependent on the [open ⇔ closed] states dynamic equilibrium of cavities between $\alpha$ and $\beta$ tubulins.

## 8 The system of librational and translational IR standing waves in the microtubules

We found out that the average length of microtubules (*l*) correlates with length of standing electromagnetic waves of librational and translational IR photons, radiated by corresponding primary effectons:

$$l_{lb} = \kappa \frac{\lambda_p^{lb}}{2} = \frac{\kappa}{2n\bar{\nu}_p^{lb}} \qquad 22$$

and

$$l_{tr} = \kappa \frac{\lambda_p^{tr}}{2} = \frac{\kappa}{2n\bar{\nu}_p^{tr}} \qquad 23$$

here $\kappa$ is the integer number; $\lambda_p^{lb,tr}$ is a librational or translational IR photon wave length equal to:

$$\lambda_p^{lb} = (n\tilde{\nu}_p^{lb})^{-1} \simeq 10^5 \text{Å} = 10\mu \qquad 24$$

$$\lambda_p^{tr} = (n\tilde{\nu}_p^{tr})^{-1} \simeq 3.5 \times 10^5 \text{Å} = 30\mu \qquad 25$$

where: $n \simeq 1.33$ is an approximate refraction index of water in the microtubule; $\tilde{\nu}_p^{lb} \simeq (700 - 750)\, cm^{-1}$ is wave number of librational photons and $\tilde{\nu}_p^{tr} \simeq (200 - 180)\, cm^{-1}$ is wave number of translational photons.

It is important that the most probable length of MTs in normal cells is about $10\mu$ indeed. So, just the librational photons and the corresponding primary effectons play the crucial role in the entanglement inside the microtubules and between MTs. The necessary for this quantum phenomena tuning of molecular dynamics of water is provided by the electromagnetic interaction between separated coherent water clusters in state of mBC.

### 8.1. The role of electromagnetic waves in the nerve cells

In the normal animal-cells, microtubules grow from pair of centriole in center to the cell's periphery. In the center of plant-cells the centrioles are absent. Two centrioles in cells of animals are always oriented at the right angle with respect to each other. The centrioles represent a construction of 9 triplets of microtubules (Fig. 52), i.e. two centriole are a source of: $(2 \times 27 = 54)$ microtubules. The centriole length is about 3000 Å and its diameter is 1000 Å.

These dimensions mean that all 27 microtubules of each centrioles can be orchestrated in the volume ($\mathbf{v}_d$) of one translational or librational electromagnetic deformon:

$$\left[\mathbf{v}_d = \frac{9}{4\pi}\lambda_p^3\right]_{tr,lb} \qquad 26$$

where: $(\lambda_p)_{lb} \sim 10^5 \text{Å}$ and $(\lambda_p)_{tr} \sim 3.5 \times 10^5 \text{Å}$

Two centrioles with normal orientation as respect to each other and a lot of microtubules, growing from them, contain the internal orchestrated system of librational water effectons. It represent a quantum system with correlated $(a \rightleftharpoons b)_{lb}^{1,2,3}$ transitions of the effectons. The resonance superradiation or absorption of a *number* of librational photons ($3q$) in the process of above transitions, is dependent on the number of primary lb effectons ($q$) in the internal hollow core of a microtubule:

$$q = \frac{\pi L_{MT}^2 \times l}{V_{ef}^{lb}} \qquad 27$$

The value of $q$ - determines the intensity (amplitude) of coherent longitudinal librational IR photons radiation from microtubule with internal radius $L_{MT} = 7nm$ and length ($l$), for the case, when condition of standing waves (22) is violated.

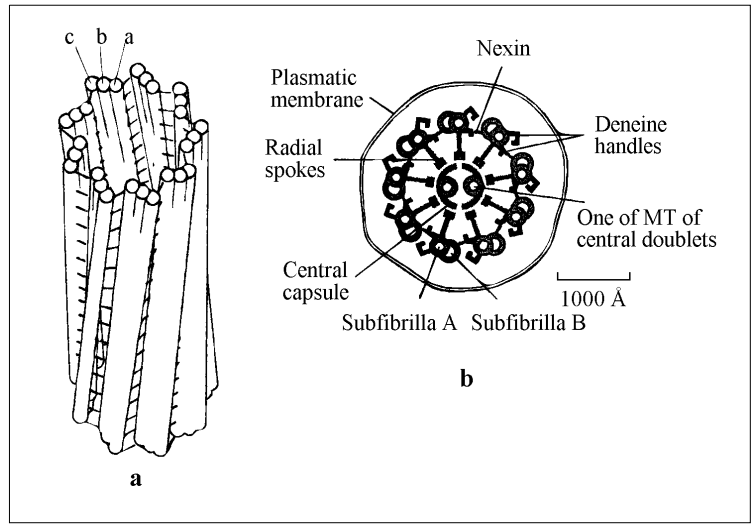

**Figure 4**. (*a*) : the scheme of centriole construction from nine triplets of microtubules. The length and diameter of cylinder are 3000 Å and 1000 Å, correspondingly. Each of triplets contain one complete microtubule and two noncomplete MT;
(b): the scheme of cross-section of cilia with number of MT doublets and MT-associated proteins (MAP): [2 × 9 + 2] = 20. One of MT of periphery doublets is complete and another is noncomplete (subfibrilles A and B).

It is important that the probabilities of pair of longitudinal and two pairs of *transversal* photons, emission as a result of superradiance by primary librational effectons are equal, being the consequence of the same collective $(b \to a)_{lb}$ transition. These probabilities can be "tuned" by the electric component of electromagnetic signals, accompanied axon polarization and nerve cell excitation due to piezoelectric properties of MT.

Coherent *longitudinal* emission of IR photons from the ends of each *pair* of microtubules of two perpendicular centrioles of the *same cell* and from ends of *one* microtubule of *other cell* can form a 3*D* superposition of standing photons (primary deformons) as a result of 3 photons pairs interception.

The system of such longitudinal electromagnetic deformons, as well as those formed by transversal photons, have a properties of pilotless 3D hologram. Such an electromagnetic hologram can be responsible for the following physico-chemical phenomena:

-Nonmonotonic distribution of intra-cell water viscosity and diffusion processes in cytoplasm due to corresponding nonmonotonic spatial distribution of macro-deformons;

-Regulation of spatial distribution of water activity $(a_{H_2O})$ in cytoplasm as a result of corresponding distribution of inorganic ions (especially bivalent such as $Ca^{2+}$) in the field of standing electromagnetic waves. Concentration of ions in the nodes of standing waves should be higher than that between them. Water activity $(a_{H_2O})$ varies in the opposite manner than ions concentration.

The spatial variation of $(a_{H_2O})$ means the modulation of [assembly ⇔ disassembly] equilibrium of filaments of the actin and partly MTs. As a consequence, the volume and shape of cell compartments will be modulated also. The activity of numerous oligomeric allosteric enzymes can be regulated by the water activity also.

The following properties of microtubules can affect the properties of 3D standing waves, radiated by them:
a) total number of microtubules in the cell;
b) spatial distribution of microtubules in the volume of cytoplasm;
c) distribution of microtubules by their length.

The constant of ($a \Leftrightarrow b$) equilibrium of primary librational effectons

$$(K_{a \Leftrightarrow b})_{lb} = \exp[-(E_a - E_b)/kT]_{lb} \qquad 28$$

and that of ($A^* \Leftrightarrow B^*$) equilibrium of super-effectons are dependent on the structure and dynamics of $\alpha\beta$ tubulin pairs forming MT walls.

This equilibrium is interrelated, in turn, with librational photons frequency $(\nu_{lb})^{1,2,3}$:

$$[\nu_{lb} = c(\tilde{\nu})_{lb} = (V_b - V_a)_{lb}/h]^{1,2,3} \qquad 29$$

which is determined by the difference of potential and total energies between (b) and (a) states of primary effectons in the hollow core of microtubules, as far the kinetic energies of these states are equal $T_b = T_a$:

$$[V_b - V_a = E_b - E_a]_{lb}^{1,2,3} \qquad 30$$

$(\tilde{\nu})_{lb}^{1,2,3}$ is the librational band wave number.

The refraction index ($n$) and dielectric constant of the internal water in MT depends on $[a \Leftrightarrow b]$ equilibrium of the effectons because the polarizability of water and their interaction in (a) state are higher, than that in (b) state.

The condition of mesoscopic Bose condensation (mBC) of any kind of liquid or solids is the increasing of librational or translational de Broglie wave length of molecules over the average separation between two neighboring molecules:

$$\lambda = \left[\frac{h}{m_{H_2O} \mathbf{v}}\right] > \left[l_m = \left(\frac{V_0}{N_0}\right)^{1/3}\right] \qquad 31$$

The dimensions and stability of water clusters in state of mBC in MTs are dependent on relative position of nonpolar cavities between alpha and beta tubulins, forming MTs and cavities dynamics.

The open state of nonpolar cavities provides the condition of *clusterphilic interaction* with water. This new kind of interaction was introduced by Kaivarainen (1985; 2000; 2007; http://arxiv.org/abs/physics/0102086) as the intermediate one between the hydrophilic and hydrophobic interactions. *The water clusters, in vicinity of nonpolar cavities are bigger and more stable, than the clusters of bulk water* (Fig. 6). As a consequence, the frequency of librational photons, radiated and absorbed in the process of quantum beats between the acoustic (a) and optic (b) states of such clusters is higher that of bulk water. Such phenomena makes the cytoplasmic water transparent for IR photons, radiated by the internal clusters of MTs. The tubulins, composing the walls of MTs are also transparent for these photons. So our approach do not consider MTs as a light-guids.

*This coherent photons exchange between remote water clusters, enhances in MTs (Fig.6) is the precondition of macroscopic Bose condensation and origination of quantum entanglement between big number of 'tuned' neurons.*

The transition of cavities between tubulins from the open to closed state of protein cavities is accompanied by water clusters disassembly and the destruction of non-uniform BC and macroscopic entanglement.

The quantum beats between the ground - acoustic (a) and excited - optic (b) states of primary librational effectons (mBC) of water are accompanied by super-radiation of coherent librational IR photons and their absorption (Kaivarainen, 1992). Similar idea for water in microtubules was proposed later by Jibu at al. (1994).

The number of coherent IR photons radiation $\rightleftharpoons$ absorption is dependent on the life-time of water cluster in the open (B) state of nonpolar cleft between α and β tubulins.

*This number can be approximately evaluated.*

The life-time of the open B- state of cleft is interrelated with that of water cluster. Our computer calculations gives a value of the life-time: $(10^{-6} - 10^{-7})$ second.

The frequency of librational IR photons, equal to frequency of quantum beats between the optical and acoustic modes of these clusters is equal to product of corresponding wave number ($\tilde{\nu}_{lb} = 700 \; cm^{-1}$) to the light velocity (c):

$$\nu_{lb} = \tilde{\nu}_{lb} c \simeq 700 \; cm^{-1} \times 3 \cdot 10^{10} \; cm/s = 2.1 \times 10^{13} s^{-1} \qquad 32$$

The characteristic time of these beats, equal to period the photons is

$$\tau = 1/\nu_{lb} \simeq 0.5 \times 10^{-13} s$$

Consequently, the number of librational photons, *radiated ⇌ absorbed* during the life-time of water cluster is:

$$n_{ph}^{lb} = \frac{\tau_{clust}}{\tau_{ph}^{lb}} = \frac{10^{-7}}{0.5 \times 10^{-13}} = 2 \times 10^6 \; photons \qquad 33$$

The corresponding photon exchange provides the EM interaction between water clusters in state of mesoscopic Bose condensation (mBC) in remote microtubules of remote neurons.

If the cumulative energy of this distant EM interaction between microtubules of centrioles, mediated by librational photons, exceeds *kT*, it may induce spatial reorientation - 'tuning' of pairs of centrioles in remote neuron's bodies.

### 9. **The reactions accompanied nerve excitation**

The normal nerve cell contains few dendrites, increasing the surface of cell's body. It is enable to form synaptic contacts for reception the information from thousands of other cells. Each neuron has one axon for transmitting the "resulting" signal in form of the electric impulses from the ends of axons of cells-transmitters to neuron-receptor.

The synaptic contacts, representing narrow gaps (about hundreds of angstrom wide) could be subdivided on two kinds: the *electric* and *chemical* ones. In chemical synapse the signal from the end of axon - is transmitted by *neuromediator, i.e. acetylholine.* The neuromediator molecules are stored in *synaptic bubbles* near *presynaptic membrane.* The releasing of mediators is stimulated by ions of $Ca^{2+}$. After diffusion throw the synaptic gap mediator form a specific complexes with receptors of post synaptic membranes on the surface of neurons body or its dendrites. Often the receptors are the ionic channels like $(Na^+, K^+)$ - ATP pump. Complex - formation of different channels with mediators opens them for one kind of ions and close for the other. Two kind of mediators interacting with channels: small molecules like acetylholine., monoamines, aminoacids and big ones like set of neuropeptides are existing..

The quite different mechanism of synaptic transmission, related to stimulation of production of secondary mediator is existing also. For example, activation of adenilatcyclase by first mediator increases the concentration of intra-cell cyclic adenozin-mono-phosphate (cAMP). In turn, cAMP can activate enzymatic phosphorylation of ionic channels, changing the electric properties of cell. This secondary mediator can participate in a lot of regulative processes, including the genes expression.

In the normal state of dynamic equilibrium the ionic concentration gradient producing by ionic pumps activity is compensated by the electric tension gradient. The *electrochemical gradient* is equal to zero at this state.

The equilibrium concentration of *Na$^+$ and Cl$^+$* in space out of cell is bigger than in cell, the gradient of *K$^+$* concentration has an opposite sign. The external concentration of very important for regulative processes *Ca$^{2+}$* (about $10^{-3}M$) is much higher than in cytosol (about $10^{-7}M$). Such a big gradient provide fast and strong increasing of *Ca$^{2+}$* internal concentration after activation of corresponding channels.

At the "rest" condition of equilibrium the resulting concentration of internal anions of neurons is bigger than that of external ones, providing the difference of potentials equal to 50-100mV. As far the thickness of membrane is only about 5nm or 50Å it means that the gradient of electric tension is about:

$$100.000 \ V/sm$$

i.e. it is extremely high.

Depolarization of membrane usually is related to penetration of *Na$^+$* ions into the cell. This process of depolarization could be inhibited by selected diffusion of *Cl$^-$* into the cell. Such diffusion can produce even *hyperpolarization* of membrane.

*The potential of action and nerve impulse can be excited in neuron - receptor only if the effect of depolarization exceeds certain threshold.*

In accordance to our model of elementary act of consciousness (EAC) three most important consequences of neuron's body polarization can occur:

- reorganization of MTs system and change of the ionic channels activity, accompanied by short-term memorization;

-reorganization of synaptic contacts on the surface of neuron and its dendrites, leading to long-term memory;

- generation of the nerve impulse, transferring the signal to another nerve cells via axons.

The propagation of nerve signal in axons may be related to intra-cellular water activity ($a_{H_2O}$) decreasing due to polarization of membrane. As a result of feedback reaction the variation of $a_{H_2O}$ induce the [*opening/closing*] of the ionic channels, thereby stimulating signal propagation along the axons.

We put forward the hypothesis, that the periodic transition of *clusterphilic* interaction of the ordered water between inter-lipid tails in nonpolar central regions of biomembranes to hydrophobic one, following by water clusters disassembly and vice verse, could be responsible for lateral nerve signal propagation/firing via axons (Kaivarainen, 1985, 1995, 2001). *The anesthetic action can be explained by its violation of the ordered water structure in the interior of axonal membranes, thus preventing the nerve signal propagation. The excessive stabilization of the internal clusters also prevent the axonal firing.*

The change of the ionic conductivity of the axonal membranes of the axons in the process of signal propagation is a secondary effect in this explanation.

The proposed mechanism, like sound propagation, can provide distant cooperative interaction between different membrane receptors on the same cell and between remote neurons bodies without strong heat producing. The latter phenomena is in total accordance with experiments.

As far the *αβ* pairs of tubulins have the properties of "electrets" (Debrabander, 1982), the *piezoelectric properties* of core of microtubules can be predicted (Athenstaedt, 1974; Mascarenhas,1974).

It means that structure and dynamics of microtubules can be regulated by electric component of electromagnetic field, which accompanied the nerve excitation. In turn, dynamics of microtubules hollow core affects the properties of internal ordered water in state of mesoscopic Bose condensation (mBC).

For example, shift of the [open ⇔ closed] states equilibrium of cavity between *α* and *β* tubulins to the open one in a course of excitation should lead to:

[I]. Increasing the dimensions and life-time of coherent clusters, represented by primary *lb* effectons (mBC)

[II]. Stimulation the distant interaction between MT of different neurons as a result of increased frequency and amplitude/coherency of IR librational photons, radiated/absorbed by primary librational effectons of internal water;

[III] Turning the mesoscopic entanglement between water molecules in coherent clusters to nonuniform macroscopic entanglement.

Twisting of the centrioles of distant interacting cells and bending of MTs can occur after [gel→sol] transition. This tuning is necessary for enhancement of the number of MTs with the parallel orientation, most effective for their remote exchange interaction by means of 3D coherent IR photons and vibro-gravitational waves.

Reorganization of actin filaments and MTs system should be accompanied by corresponding changes of neuron's body and its dendrites shape and activity of certain ionic channels and synaptic contacts redistribution; This stage is responsible for long-term memory emergency.

At [sol]-state the $Ca^{2+}$ - dependent $K^+$ channels turns to the open state and internal concentration of potassium decreases. The latter oppose the depolarization and decrease the response of neuron to external stimuli. Decay of neuron's response is termed "adaptation". This *response adaptation* is accompanied by *MTs-adaptation*, i.e. their reassembly in conditions, when concentration of $Ca^{2+}$ tends to minimum. The reverse [sol→gel] transition stabilize the new equilibrium state of given group of cells.

The described hierarchic sequence of stages: from mesoscopic Bose condensation to macroscopic one, providing entanglement of big number of cells, their simultaneous synaptic reorganization and synhronization of the excitation ⇌ relaxation cycles of nerve cells, are different stages of elementary act of consciousness.

## 10. Possible mechanism of wave function collapsing, following from the Elementary Act of Consciousness

A huge number of superimposed possible quantum states of any quantum system always turn to "collapsed" or "reduced" single state as a result of measurement, i.e. interaction with detector.

In accordance to "Copenhagen interpretation", the collapsing of such system to one of possible states is unpredictable and purely random. Roger Penrose supposed (1989) that this process is due to quantum gravity, because the latter influences the quantum realm acting on space-time. After certain gravity threshold the system's wave function collapsed "under its own weight".

Penrose (1989, 1994) considered the possible role of quantum superposition and wave function collapsing in synaptic plasticity. He characterized the situation of learning and memory by synaptic plasticity in which neuronal connections are rapidly formed, activated or deactivated: "Thus not just one of the possible alternative arrangements is tried out, but vast numbers, all superposed in complex linear superposition". The collapse of many cytoskeleton configuration to single one is a nonlocal process, required for consciousness.

This idea is in-line with our model of elementary acts of consciousness as a result of transitions between nonuniform macroscopic and mesoscopic Bose condensation (BC) of big number of electromagnetically tuned neurons and corresponding oscillation between their entangled and non-entangled states.

Herbert (1993) estimated the mass threshold of wave function collapse roughly as $10^6$

daltons. Penrose and Hameroff (1995) calculated this threshold as

$$\Delta M_{col} \sim 10^{19} D \qquad 34$$

Non-computable self-collapse of a quantum coherent wave function within the brain may fulfill the role of non-deterministic free will after Penrose and Hameroff (1995).

For the other hand, in accordance with proposed in this author model, the induced coherency between coherent water clusters (primary librational effectons - mesoscopic Bose condensate) in MTs, as a result of distant exchange of librational photons, emitted ⇌ absorbed by them, leads to formation of *macroscopic BC* in microtubules.

The increasing of the total mass of water, involved in macroscopic nonuniform BC in a big system of remote MTs and corresponding 'tuned' neuron ensembles, up to gravitational threshold may induce the wave function collapse in accordance to Penrose hypothesis.

In our approach we explain the selection of certain configurational space of huge number of 'tuned' neurons, not by structural changes of tubulins like in Hameroff-Penrose model, but by increasing of mass of water in state of macroscopic BC in brain in the process of condensation of spatially separated mesoscopic BC (coherent water clusters in MTs). The macroscopic BC is initiated by correlated shift of dynamic equilibrium ($a \rightleftharpoons b$) of nonpolar cavities, formed by pairs of tubulins, between the open ($b$) and closed ($a$) states to the open one, stabilizing water clusters. The time of development/evolution of coherence in remote neurons, accompanied by increasing of scale of macroscopic BC is much longer, than that of mesoscopic BC (about $10^{-6}$ s, equal to average period of pulsation of tubulin dimers cavity between open and close state) and can be comparable with time between axonal firing (about $1/40 = 2.5 \times 10^{-2}$ s). The time of coherence determines the period between corresponding wave function collapsing.

The corresponding structural rearrangements of tubulins and their pairs in the process of shift of open⇌ closed clefts to the right or left, do not change their mass and can not be a source of wave function collapsing "under its own weight" in contrast to increasing of mass of water in evolution of nonuniform macroscopic BC from mBC.

The dynamics of $[\text{increasing} \rightleftharpoons \text{decreasing}]$ of the entangled water mass in state of macroscopic BC is a result of correlated shift of dynamic equilibrium between primary *librational (lb) effectons* (coherent water clusters, mBC), stabilized by *open* inter-tubulins cavities and primary *translational (tr) effectons (independent water molecules)*, corresponding to closed cavities.

The correlated conversions between librational and translational effectons $[lb \rightleftharpoons tr]$ of water in remote MTs, representing the association ⇌ dissociation of the entangled water clusters in state of mBC reflect, in fact, the reversible cycles of [coherence ⇌ decoherence] corresponding to cycles of mesoscopic wave function of these clusters collapsing. The relatively slow oscillations of dynamic equilibrium of $[lb \rightleftharpoons tr]$ conversions are responsible for alternating contribution of macroscopic quantum entanglement in consciousness.

*Let's make some simple quantitative evaluations in proof of our interpretation of the wave function collapsing.* The mass of water in one microtubule in nerve cell body with most probable length $\sim 10^5$ Å and diameter 140 Å is about

$$m_{H_2O} \sim 10^8 D$$

In accordance with our calculations for bulk water, the fraction of molecules in composition of primary *tr* effectons is about 23% and that in composition of primary *librational* effectons (mBC) is about ten times less (Figure 53) or 2.5%. In MTs due to clusterphilic interaction, stabilizing water clusters, this fraction mBC can be few times bigger.

We assume, that in MTs at least 10% of the total water mass ($10^8 D$) can be converted to primary librational effectons (coherent clusters) as a result of IR photons exchange and entanglement between mBC of the same MTs. This corresponds to increasing of mass of these quasiparticles in each MT as:

$$\Delta m_{H_2O} \simeq 10^6 D \qquad 35$$

Such increasing of the coherent water fraction is accompanied by decreasing of water mass, involved in other types of excitations in MT.

Based on known experimental data that each nerve cell contains about 50 microtubules, we assume that the maximum increasing of mass of primary librational effectons in one cell, using 35, could be:

$$\Delta M_{H_2O} \sim 50\, \Delta m_{H_2O} = 5 \times 10^7 D \qquad 36$$

If the true value of mass threshold, responsible for wave function collapse, $\Delta M_{col}$ is known (for example $10^{16} D$), then the number ($N_{col}$) of neurons in assemblies, required for this process is

$$N_{col} \sim (\Delta M_{col}/\Delta M_{H_2O}) = 2 \times 10^8 \qquad 37$$

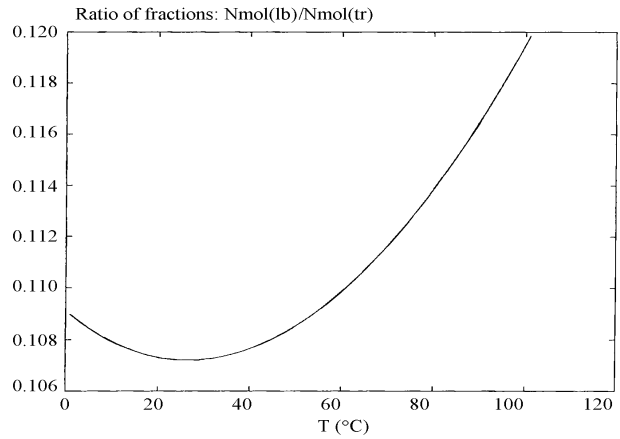

**Figure 5**. Calculated ratio of water fractions involved in primary [lb] effectons to that, involved in primary [tr] effectons for the bulk water.

MAP– microtubules associated proteins stabilize the overall structure of MTs. They prevent the disassembly of MTs in bundles of axons and *cilia* in a course of their coherent bending. In neuron's body the concentration of MAP and their role in stabilization of MTs is much lower than in cilia.

The total number of nerve cells in human brain is about: $N_{tot} \sim 10^{11}$. The critical fraction of cells population, participating in elementary act of consciousness, following from our model, can be calculated as:

$$f_c = (N_{col}/N_{tot}) \sim 0.05 \qquad 38$$

This value is dependent on correct evaluation of critical mass ($\Delta M_{col}$) of collapsing.

The [gel $\to$ sol] transition, induced by coherent "collapsing" of macroscopic brain wave function and dissociation of water clusters in MTs in state BC of huge number of tuned neurons, followed by synaptic contacts reorganization, represents the elementary act of consciousness.

Our approach agree with general idea of Marshall (1989) that Bose- condensation could be responsible for "unity of conscious experience". However, our model explains how this idea can work in detail and what kind of Bose condensation is necessary for its realization.

*We can resume now, that in accordance with our Elementary Act of Consciousness, the sequence of following interrelated stages is necessary for elementary act of perception and memory:*

1. The change of the electric component of cell's electromagnetic field as a result of neuron depolarization;

2. Shift of $A \rightleftharpoons B$ equilibrium between the closed (A) and open to water (B) states of cleft, formed by $\alpha$ and $\beta$ tubulins in microtubules (MT) to the right due to the piezoelectric effect;

3. Increasing the life-time and dimensions of coherent "flickering" water clusters, representing the 3D superposition of de Broglie standing waves of $H_2 0$ molecules with properties of Bose-condensate (*effectons*) in hollow core of MT. This process is stimulated by the open nonpolar clefts between tubulin dimers in MT;

4. Spatial "tuning" of parallel MTs of distant simultaneously excited neurons due to distant electromagnetic interaction between them by means of superradiated IR photons and centrioles twisting;

5. Turning the mesoscopic BC of $H_2O$ molecules in clusters to nonuniform macroscopic BC, mediated by Virtual Replica of the clusters Multiplication in space VRM(r) and accompanied by activation of nonlocal interaction between remote clusters in big number of entangled MTs of neurons dendrites;

6. Destabilization of superimposed wave function eigenvalues of clusters (mBC) as a result of nonlinear optical effects like *bistability and self-induced transparency and superradiation;*

7. Disassembly of the actin filaments and [gel-sol] transition, decreasing strongly both - the viscosity of cytoplasm and water activity;

8. The coherent volume/shape pulsation of big group of interacting cells as a consequence *of (actin filaments+MTs) system disassembly and water activity decreasing.* The latter occur as a result of increasing of water fraction in hydration shell of actin and tubulin subunits due to increasing of their surface after disassembly. The decreasing of cytoplasmic water activity increases the passive osmoses of water from the external volume to the cell, increasing its volume.

*This stage should be accompanied by four effects:*

*(a)* Increasing the volume of the nerve cell body;

*(b)* Disrupting the (+) ends of MTs with cytoplasmic membranes, making MTs possible to bend in cell and to collective spatial tuning of huge number of MTs in the ensembles of even distant excited neurons;

(c) Origination of new MTs and microtubules associated proteins (MAP) system switch on/off the ionic channels and change the number and properties of synaptic contacts, responsible for short (MAP) and long memory;

(d) Decreasing the concentration of $Ca^{2+}$ to the limit one, when its ability to disassembly of actin filaments and MT is stopped and [gel $\rightleftharpoons$ sol] equilibrium shifts to the left again, stabilizing a new MTs and synaptic configuration.

This cyclic consequence of quantum mechanical, physico-chemical and nonlinear classical events can be considered as elementary act of memory and consciousness realization. This act can be as long as 500 ms, *i.e.* half of second, like proposed in Hamroff-Penrose model.

The elementary act of consciousness include a stage of coherent electric firing in brain (Singer, 1993) of distant neurons groups with period of about 1/40 sec.

Probability of super-deformons and cavitational fluctuations increases after [gel→sol] transition. This process is accompanied by high-frequency (UV and visible) "biophotons" radiation due to recombination of part of water molecules, dissociated as a result of cavitational fluctuation.

The dimension of IR super-deformon edge is determined by the length of librational IR standing photon - about 10 microns. It is important that this dimension corresponds to the average microtubule length in cells confirming in such a way our idea. Another evidence in proof is that is that the resonance wave number of excitation of super-deformons, leading from our model is equal to 1200 $(1/cm)$.

The experiments of Albreht-Buehner (1991) revealed that just around this frequency the response of surface extensions of 3T3 cells to weak IR irradiation is maximum. Our model predicts that IR irradiation of microtubules system *in vitro* with this frequency will dramatically increase the probability of gel →sol transition.

*Except super-radiance, two other cooperative optic effects could be involved in supercatastrophe realization: self-induced bistability and the pike regime of IR photons radiation (Bates, 1978; Andreev et al.,1988).*

The characteristic frequency of pike regime can be correlated with frequency of [gel-sol] transitions of neuronal groups in the head brain.

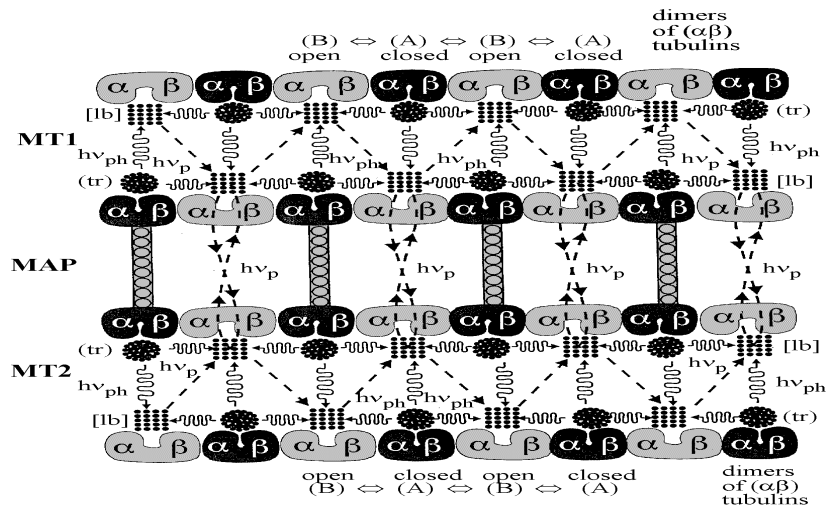

**Figure 6**. The correlation between local, conformational and distant - electromagnetic interactions between pairs of tubulins and microtubules (MT1 and MT2), connected by MAP by mean of librational IR photons exchange.

The dynamics of [ *increasing* ⇌ *decreasing* ] of the entangled water mass in state of macroscopic BC in the process of elementary act of consciousness is a result of correlated shift of dynamic equilibrium between open and closed cavities between alpha and beta tubulins. As a result of these cavities transition from the open to closed state the primary *librational (lb) effectons* (coherent water clusters in state of mesoscopic Bose condensation - mBC) disassembly to small primary *translational (tr) effectons* (independent water molecules), induced by transition of the open states of cavities to the closed one. The nonuniform macroscopic entanglement between the remote water clusters in state of mBC is stimulated by coherent IR photons exchange and vibro-gravitational interaction between these clusters.

MAP– microtubules associated proteins stabilize the overall structure of MTs. They prevent the disassembly of MTs in bundles of axons and *cilia* in a course of their coherent bending. In neuron's body the concentration of MAP and their role in stabilization of MTs is much lower than in cilia (Kaivarainen, 1995, 2003).

The distant electromagnetic and vibro-gravitational interactions between different MT are the consequence of IR photons and coherent vibro-gravitational waves exchange. The corresponding two types of waves are excited as a result of correlated ($a \Leftrightarrow b$) transitions of water primary librational effectons, localized in the open B- states of ($\alpha\beta$) clefts. Frequency of ($a \Leftrightarrow b$) transitions and corresponding superradiated IR photons is about $2 \times 10^{13}\ s^{-1}$. It is much higher, than frequency of transitions of clefts of $\alpha\beta$ tubulin dimers between open and closed states.

When the neighboring ($\alpha\beta$) clefts has the alternative open and closed states like on Fig 54, the general spatial structure remains straight. However, when [$A \Leftrightarrow B$] equilibrium of all the clefts from one side of MT are shifted to the left and that from the opposite side are shifted to the right, it leads to bending of MT. Coherent bending of MTs could be responsible for [volume/shape] vibrations of the nerve cells and the cilia bending.

Max Tegmark (2000) made evaluation of decoherence time of neurons and microtubules for analyzing the correctness of Hameroff-Penrose idea of wave function collapsing as a trigger of neurons ensembles axonal firing.

The following three sources of decoherence for the ions in the act of transition of neuron between 'resting' and 'firing' are most effective:
1. Collisions with other ions
2. Collisions with water molecules
3. Coulomb interactions with more distant ions.

The coherence time of such process, calculated from this simple model appears to very short: about $10^{-20}$ s.

The electrical excitations in tubulins of microtubules, which Penrose and others have suggested may be relevant to consciousness also where analyzed. Tegmark considered a simple model of two separated but superimposed (entangled) positions of kink, travelling along the MT with speed higher than 1 m/s, as it supposed in Hameroff-Penrose (H-P) model. The life-time of such quantum state was evaluated as a result of long-range electromagnetic interaction of nearest ion with kink.

His conclusion is that the role of quantum effects and wave function collapsing in H-P model is negligible because of very short time of coherence: $10^{-13}$ s for microtubules.

Hagan, Hameroff and Tuszynski (2002) responded to this criticism, using the same formalism and kink model. Using corrected parameters they get the increasing of life-time of coherent superposition in H-P model for many orders, up to $10^{-4}$ s. This fits the model much better.

Anyway the approach, used by Tegmark for evaluation of the time of coherence/entanglement is not applicable to our model, based on oscillation between mesoscopic and macroscopic nonuniform Bose condensation. It follows from our approach that even very short life-time of oscillating semivirtual macroscopic Bose condensation can be effective for realization of entanglement.

## 11. Experimental data, confirming the Hierarchic Model of Consciousness

There are some experimental data, which support the role of microtubules in the information processing. Good correlation was found between the learning, memory peak and intensity of tubulin biosynthesis in the baby chick brain (Mileusnic *et al.*1980). When baby rats begin their visual learning after they first open eyes, neurons in the visual cortex start to produce vast quantities of tubulin (Cronley- Dillon et. al., 1974). Sensory stimulation of goldfish leads to structural changes in cytoskeleton of their brain neurons (Moshkov *et al.*, 1992).

There is evidence for interrelation between cytoskeleton properties and nerve membrane excitability and synaptic transmission (Matsumoto and Sakai, 1979; Hirokawa, 1991). It has been shown, that microtubules can transmit electromagnetic signals between membranes (Vassilev *et al.*, 1985).

Desmond and Levy (1988) found out the learning-associated change in dendritic spine shape due to reorganization of actin and microtubules containing, cytoskeleton system. After "learning" the number of receptors increases and cytoskeleton becomes more dense.

Other data suggest that cytoskeleton regulates the genome and that signaling along microtubules occurs as cascades of phosphorylation/dephosphorylation linked to calcium ion flux (Puck, 1987; Haag et al, 1994).

The frequency of super-deformons excitations in bulk water at physiological temperature ($37^0C$) is around:

$$\nu_S = 3 \times 10^4 \ s^{-1}$$

The frequency of such cavitational fluctuations of water in MT, stimulating in accordance to our model cooperative disassembly of the actin and partly MT filament, accompanied by gel→ *sol* transition, could differ a bit from the above value for bulk water.

Our model predicts that if the neurons or other cells, containing MTs, will be treated by acoustic or electromagnetic field with resonance frequency of intra-MT water ($\nu_{res} \sim \nu_S^{MT} \geq 10^4 s^{-1}$), it can induce simultaneous disassembly of the actin filaments and destabilization of MTs system, responsible for maintaining the specific cell volume and geometry. As a result, it activates the neuron's body volume/shape pulsation.

Such external stimulation of *gel* ⇌ *sol* oscillations has two important consequences:

*-The first one* is generation of oscillating high-frequency nerve impulse, propagating via axons and exciting huge number of other nerve cells, i.e. distant nerve signal transmission in living organism;

*-The second one* is stimulation the leaning process as far long-term memory in accordance to Elementary Act of Consciousness, is related to synaptic contacts reorganization, accompanied the neuron volume/shape pulsation.

Lehardt *et al.* (1989) supposed that ultrasonic vibrations are perceptive by tiny gland in the inner ear, known as the *Saccule*. It looks that *Saccule may* have a dual functions of detection *gravity and auditory signals. Cohlea could be a result of Saccule evolution in mammals.*

Lenhardt and colleagues constructed the an amplitude modulated by audio-frequencies ultrasonic transmitter that operated at frequencies: (28-90) kHz. The output signal from their device was attached to the deaf people heads by means of piezoelectric ceramic vibrator. All people "heard" the modulated signal with clarity.

Our approach allows to predict the important consequence. Excitation of water super-deformons in cells, leading to gel→ *sol* transition, cell's volume/shape pulsation and generation of high-frequency nerve impulse - could be stimulated by hypersound, electromagnetic waves and coherent IR photons with frequency, corresponding to excitation energy of super-deformons.

The calculated from this assumption frequency is equal to

$$\nu_p^S = c \times \widetilde{\nu}_p^S = (3 \times 10^{10}) \times 1200 = 3.6 \times 10^{13} \, c^{-1} \qquad 39$$

the corresponding photons wave length:

$$\lambda_p^S = c/(n_{H_2O} \times \nu_p^S) \simeq 6.3 \times 10^{-4} \, cm = 6.3 \mu \qquad 40$$

where $\tilde{\nu}_p^S = 1200\ cm^{-1}$ is wave number, corresponding to energy of super-deformons excitation;

$n_{H_2O} \simeq 1.33$ is refraction index of water.

### *11.1 The additional experimental verification of the Hierarchic Model of Consciousness "in vitro"*

It is possible to suggest some experimental ways of verification of Elementary Act of Consciousness using model systems. The important point of Elementary Act of Consciousness is stabilization of highly ordered water clusters (primary librational effectons) in the hollow core of microtubules. One can predict that in this case the IR librational bands of water in the oscillatory spectra of model system, containing sufficiently high concentration of MTs, must differ from IR spectra of bulk water as follows:

- the shape of librational band of water in the former case must contain 2 components: the first one, big and broad, like in bulk water and the second one small and sharp, due to increasing coherent fraction of librational effectons. The second peak should disappeared after disassembly of MTs with specific reagents;

- the velocity of sound in the system of microtubules must be bigger, than that in disassembled system of MTs and bulk pure water due to bigger fraction of ordered ice-like water;

- all the above mentioned parameters must be dependent on the applied electric potential, due to piezoelectric properties of MT;

- the irradiation of MTs system in vitro by ultrasonic or electromagnetic fields with frequency of super-deformons excitation of the internal water of MTs at physiological temperatures $(25 - 40^0 C)$ :

$$\nu_s = (2 - 4) \times 10^4\ Hz$$

have to lead to increasing the probability of disassembly of MTs, induced by cavitational fluctuations. The corresponding effect of decreasing turbidity of MT-containing system could be registered by light scattering method.

Another consequence of super-deformons stimulation by external fields could be the increasing of intensity of radiation in visible and UV region due to emission of corresponding "biophotons" as a result of recombination reaction of water molecules:

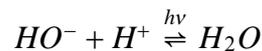

$$HO^- + H^+ \overset{h\nu}{\rightleftharpoons} H_2O$$

Cavitational fluctuations of water, representing in accordance to our theory super-deformons excitations, are responsible for dissociation of water molecules, *i.e.* elevation of protons and hydroxyls concentration. These processes are directly related to sonoluminiscence phenomena.

The coherent transitions of $(\alpha\beta)$ dimers, composing MTs, between "closed" (A) and "open" (B) conformers with frequency $(\nu_{mc} \sim 10^7\ s^{-1})$ are determined by frequency of water macroconvertons (flickering clusters) excitation, localized in cavity between $\alpha$ and $\beta$ tubulins. If the charges of (A) and (B) conformers differ from each other, then the coherent $(A \rightleftharpoons B)$ transitions generate the vibro-gravitational and electromagnetic field with the same radio-frequency. The latter component of biofield could be detected by corresponding radio waves receiver.

We can conclude that the Hierarchic Theory of condensed matter and its application to water and biosystems provide reliable models of informational exchange between different cells and correlation of their functions. Hierarchic Model of consciousness is based on proposed quantum exchange mechanism of interactions between neurons, based on very special properties of microtubules, [gel-sol] transitions and interrelation between spatial distribution of MTs in neurons body and synaptic contacts.

The described mechanism of IR photons - mediated conversion of mesoscopic Bose condensation to macroscopic one with corresponding change of wave function spatial scale can be exploit in the construction of artificial quantum computers, using inorganic microtubules.

===============================================================================

**Appendix**

The entropy-driven information processing

It follows from our Elementary Act of Consciousness or Cycle of Mind that changes of system of electromagnetic, acoustic and vibro-gravitational 3D standing waves, stimulating interconversions between mesoscopic and macroscopic Bose condensation of the internal water of MTs change the informational content of this water.

This process induces redistribution of probabilities of different water excitations in huge number of microtubules. It means corresponding change of informational entropy $<I>$, related to microtubules in accordance with known relations (Kaivarainen 1995; 2007; http://arXiv.org/abs/physics/0003107):

$$<I> = \sum_i P_i \lg(1/P_i) = -\sum_i P_i \lg(P_i) \qquad \text{A.1}$$

where: $P_i$ is a probability of the ($i$) state (excitation) with energy ($E_i$), defined as:

$$P_i = \frac{\exp(-\frac{E_i}{kT})}{\sum_i \exp(-\frac{E_i}{kT})} \qquad \text{A.2}$$

For the total system the well known relation between entropy (S) and information (I) is:

$$S(e.u.) = k \ln W = (k \ln 2)I = 2.3 \times 10^{-24} I \ (bit) \qquad \text{A.3}$$

where statistical weight of macrosystem:

$$W = \frac{N!}{N_1! N_2! \ldots N_i!} \qquad \text{A.4}$$

the total number of internal water molecules in macrosystem of interacting MT is:
$N = N_1 + N_2 + \ldots + N_i$;

[$i$] is number of non degenerated states of 24 quasiparticles of intra MT water.

The information amplitude of condensed matter [http://arXiv.org/abs/physics/0003107] we introduce as a product of the number of molecules ($N_i$) in each of [$i$] collective excitations:

$$N_m^i = \mathbf{V}_i / \mathbf{v}_{H_2O} = (1/n_i)/(V_0/N_0) \qquad \text{A5}$$

and informational entropy (A.1):

$$< \mathbf{IA} > \ = \ < I > N_m^i = - \frac{N_0}{V_0} \sum_i \frac{P_i}{n_i} \lg_2(P_i) \qquad \text{A.6}$$

where $N_0$ and $V_0$ are the Avogadro number and molar volume of water; $\mathbf{V}_i = 1/n_i$ is a volume of excitation of ($i$)-type; $n_i$ is concentration of corresponding excitations.

The distant IR photons exchange between water in MT and oscillation between mesoscopic and nonuniform macroscopic Bose condensation, accompanied by the change of probability $P_i$ of different excitations can be considered as an informational exchange between nerve cells. It is accompanied by change of fractions of water excitations in system of interacting MTs, providing conditions for macroscopic entanglement between coherent water clusters.